\documentclass[12pt,a4paper]{article}
\usepackage[utf8]{inputenc}
\usepackage{amsmath}
\usepackage{amsfonts}
\usepackage{amssymb}
\usepackage{blindtext}
\usepackage{authblk}
\usepackage{fancyhdr}
\usepackage{hyperref}
\usepackage{graphicx}
\usepackage{lmodern}
\usepackage[left=2cm,right=2cm,top=2cm,bottom=2cm]{geometry}
\begin{document}

\title{Neutron Proton Scattering for S, P and D Waves using Deng-Fan Potential by Phase Function Method}
\author[1]{Ayushi Awasthi \thanks{awasthiayushi1998@gmail.com}}
\author[1]{O.S.K.S. Sastri  \thanks{Corresponding Author: sastri.osks@hpcu.ac.in}}
\affil[1]{Department of Physics and Astronomical Science, Central University of Himachal Pradesh, Dharmshala , 176215}

\maketitle
\begin{abstract}
In this paper, the np - scattering phase shifts and cross section for S,P and D partial waves have been obtained for energies below the pion threshold, by considering Deng-Fan potential as model of interaction. The radial time independent Schrödinger equation has been analytically solved using Nikiforov - Uvarov method to obtain the energy expression for ground state of np system. Utilising this, the  scattering phase shifts for $^3S_1$ have been obtained using phase function method. The phase equations for various scattering states $1S_0$, $^1P_1$, $^3P_{0,1,2}$, $^1D_2$, and $^3D_{1,2,3}$ have been numerically solved for obtaining corresponding scattering phase shifts and their respective partial cross section. The total scattering cross sections computed at various energies are found to be closely matching with experimental data. The low energy scattering parameters determined from scattering phase shifts of $^3S_1$ and $^1S_0$ are reasonably close to experimental ones. Hence, Deng-Fan potential is a good phenomenological potential to understand the np - scattering system.


\end{abstract} 
\section{Introduction}\label{sec1}

The study of the nucleon-nucleon (NN) interaction has indeed been a central and recurring problem in nuclear physics for decades. The NN interaction is the fundamental force that governs the behavior of protons and neutrons  within atomic nuclei. Understanding this interaction is crucial for explaining the properties and behaviors of atomic nuclei, as well as for shedding light on the underlying nature of the nuclear force \cite{Machleidt}.
One of the important goals of nuclear physics is to model the interaction, by a phenomenological potential, responsible for explaining observed scattering cross-sections (SCS) by phase shift analysis or phase wave analysis \cite{Hans}. This involves solving the non-relativistic radial time independent Schrödinger equation (TISE) for the chosen potential for various $\ell$-channels, called as partial waves, to obtain corresponding wavefunctions. One deduces  scattering phase shifts (SPS) by matching wavefunction within potential region with asymptotic solution \cite{Schiavilla}. These SPS are utilised to calculate the partial and total SCS. The partial SCS would give information about resonances. The match between experimental resonances and total cross sections with those obtained by theoretical interaction potential validate the model. While most of the techniques like R-matrix \cite{Wigner}, Jost function method (JFM) \cite{Jost}, complex scaling method (CSM) \cite{Odsuren}, J-matrix method \cite{Alhaidari}, etc., are dependent on wavefunction, variable phase approach (VPA) or phase function method (PFM) \cite{Zhaba,Calogero,Babikov} directly obtains SPS from interaction potential. The phase function method
turns out to be more efficient and has the ability to resolve a
much smaller phase shift than the Schrödinger equation
method, in certain regions of the parameter space \cite{Xu}. Our group has successfully utilised PFM for studying various two body nuclear scattering\cite{Sastri,Khachi}.\\
Yukawa\cite{Yukawa} gave the first successful model of interaction for np system. Using PFM, we have found scattering phase shift(SPS) for Yukawa potential to be having good match with expected values for energies upto 50 MeV only\cite{Pérez}. To explain the expected SPS for energies below pion-threshold, considered to be 350 MeV, a repulsive core was added by Malfliet-Tjon (MT)\cite{MT}, with reasonable success \cite{Shikha}. Both these phenomenological potentials had the disadvantage of not having analytical solutions for their corresponding radial time independent Schrödinger equation (TISE). Recently, an analytical solution for Yukawa had been obtained using Super Symmetric (SUSY) technique \cite{Napsuciale}. Originally, Hulthen\cite{Hulthen} suggested a variation of Yukawa and obtained a simple analytical solution, due to which much progress was made in understanding the Deuteron wavefunction. On observing the choice of repulsive term in MT potential, one is intrigued to find that a similar form deduced for Hulthen would lead to one of the simplest Deng-Fan (DF) potential \cite{Deng}, which is chosen as model of interaction for this work. This DF potential has been used to describe binary structure in $^{19}Ne$ nuclei \cite{binary}. The DF potential belonging to the class of exponential type functions, has the advantage of having acceptable physical boundary conditions at r=0 $\&$ $\infty$ \cite{Mesa}. Recently, Saha et. al. \cite{Saha} used DF potential to obtain SPS and cross sections for np and nd system, by both Jost function and phase function method. Their study of np scattering was limited to data up to 50 MeV only and hence did not utilize the repulsive nature in-built into DF potential which could explain negative SPS at higher energies. On the other hand, we have utilised Morse function, which can be considered to be limiting case of DF potential, for studying np scattering \cite{Physica} and structure of deuteron \cite{PRC} very comprehensively.
\\ In this paper, our goal is to investigate the effectiveness of DF potential for explaining the experimentally obtained np scattering cross sections for energies all the way up to 350 MeV, by considering S, P and D channels.  
\section{Theoretical Framework}
The Deng-Fan potential is given by \cite{Saha}:
\begin{equation}
V_{DF}(r) = - V_{1} \frac{e^{-\alpha r}}{(1- e^{-\alpha r})} + V_{2} \frac{e^{-2 \alpha r}}{(1- e^{-\alpha r})^{2}}
\label{DF}
\end{equation}
where $V_1$ and $V_2$ represent the strengths of the potential measured in units of MeV, and $\alpha$ denotes the inverse range parameter with dimensions of $fm^{-1}$.
The radial Schrödinger equation for the potential function defined in eq. \ref{DF} takes the form when considering $\ell$ = 0:
\begin{equation}
\frac{d^{2}u(r)}{dr^{2}} + \frac{2\mu}{\hbar^2}\left[E - \left(- V_{1} \frac{e^{-\alpha r}}{(1- e^{-\alpha r})} + V_{2} \frac{e^{-2 \alpha r}}{(1- e^{-\alpha r})^{2}}\right)\right]u(r) =0
\label{SE}
\end{equation}
To obtain the exact or approximate solutions of the Schrödinger equation for an exponential-type potential, various methods have been employed. These methods include the Nikiforov-Uvarov method (NU)\cite{NU,NU1}, the factorization method \cite{Dong,Infeld}, the asymptotic iteration method \cite{Ozer,Ciftci}, and others \cite{Davydov}. We are employing the NU method to solve the radial TISE for the Deng-Fan potential to determine its vibrational states.
\subsection{Nikiforov-Uvarov (NU) method}
The Schrödinger equation in spherical coordinates is reduced for a given potential to a generalized equation of hypergeometric type with an appropriate coordinate transformation, r $\rightarrow$ s, as \cite{NU,NU1}
\begin{equation}
\frac{d^{2}\psi(s)}{ds^{2}} +\frac{\tilde{\tau}(s)}{\sigma(s)}\frac{d\psi(s)}{ds}+\frac{\tilde{\sigma}(s)}{\sigma^{2}(s)}\psi(s)=0
\label{NU}
\end{equation}
where  $\sigma(s)$ and $\tilde{\sigma(s)}$ are represented as polynomials with a maximum degree of two, while $\tilde{\tau}(s)$ takes the form of a first-degree polynomial. Furthermore, the function $\psi(s)$ can be expressed as a hypergeometric-type function. Upon applying the transformations denoted by $\psi(s) = \psi(s)y(s)$ to equation \ref{NU}, a resulting set of differential equations emerges as follows:
\begin{equation}
\sigma(s)y^{\prime \prime}(s) + \tau(s)y^{\prime}(s)+\lambda y(s) =0
\end{equation}
\begin{equation}
\frac{d}{ds}[ln\phi(s)]=\frac{\pi(s)}{\sigma(s)};~~~~~\psi(s) = \frac{1}{2}[\tau(s)-\tilde{\tau}(s)]
\end{equation}
where the polynomial solutions $y_n(s)$ are given by Rodrigues’ formula
\begin{equation}
y_n(s) = \frac{B_n}{\rho(s)}\frac{d^{n}}{ds^{n}}[\sigma^{n}(s)\rho(s)]
\end{equation}
with the normalized constant $B_{n}$ and the weight function $\rho(s)$ satisfying a equation
\begin{equation}
\frac{d}{ds}[\sigma^{n}(s)\rho(s)]=\tau(s)\rho(s)
\end{equation}
The function $\pi(s)$ and the parameter $\lambda$ are defined as
\begin{equation}
\pi(s) = \frac{\sigma^{\prime}(s)-\tilde{\tau}(s)}{2}\pm \sqrt{\left(\frac{\sigma^{\prime}(s)-\tilde{\tau}(s)}{2}\right)^{2}-\tilde{\sigma}(s)+k\sigma(s)}
\label{pi}
\end{equation}
where 
\begin{equation}
\lambda = k+\pi^{\prime}(s)
\label{1}
\end{equation}
The discriminant of the expression under the square
root in the polynomial $\pi(s)$ must be zero, which defines the constant k. Thus a new eigenvalue equation becomes
\begin{equation}
\lambda_{n}=-n\tau^{\prime}(s)-\frac{1}{2}n(n-1)\sigma^{\prime \prime}(s) ~, ~~~~~~n = 0,1,2.....
\label{2}
\end{equation}
where the derivative of $\tau(s) = \tilde{\tau}(s)+2\pi(s)$ must be
negative.
\\To convert the eq.\ref{SE} into hyper-geometric-type
second-order differential equations, consider the transformation of the form $e^{-\alpha r} = s$ and $ u(r) \rightarrow \psi(s)$ and re-write eq.\ref{SE} as
\begin{equation}
\frac{d^{2}\psi(s)}{ds^{2}} +\frac{(1-s)}{s(1-s)}\frac{d\psi(s)}{ds}+\frac{2\mu}{\hbar^2\alpha^2s^{2}}\left[(E-\frac{V_1 s}{(1-s)}-\frac{V_2 s^2}{(1-s)^2}\right]\psi(s) = 0
\label{HE}
\end{equation}
\begin{equation}
\frac{d^{2}\psi(s)}{ds^{2}} +\frac{(1-s)}{s(1-s)}\frac{d\psi(s)}{ds}+\frac{2\mu}{\hbar^2\alpha^2s^{2}(1-s)^2}\left[(E-V_1-V_2)s^2 +(V_1-2E)s +E\right]\psi(s) =0
\end{equation}

By introducing the following dimensionless parameters
\begin{equation}\nonumber
\epsilon^{2} = -\frac{2\mu E}{\hbar^2\alpha^2} ;~~ 
\beta_1 = \frac{2\mu V_1}{\hbar^2\alpha^2} ; ~~
\beta_2 = \frac{2\mu V_2}{\hbar^2\alpha^2}
\end{equation}
which leads to the equation
\begin{equation}
\frac{d^{2}\psi(s)}{ds^{2}} +\frac{(1-s)}{s(1-s)}\frac{d\psi(s)}{ds}+\frac{1}{s^{2}(1-s)^2}\left[-(\epsilon^2+\beta_1+\beta_2)s^2 +(2\epsilon^2+\beta_1)s - \epsilon^2\right]\psi(s) =0
\label{HE1}
\end{equation}
\\On comparing eq.\ref{HE1} with eq. \ref{NU}, we get 
\begin{center}
$\tilde{\tau}(s)$ = (1-s); $\sigma(s)$ = s~(1-s);  
$\tilde{\sigma}(s) = -(\epsilon^2+\beta_1+\beta_2)s^2 +(2\epsilon^2+\beta_1)s - \epsilon^2$
\end{center}
On substituting these polynomials into eq. \ref{pi}, we obtain the polynomial $\pi(s)$
\begin{equation}
\pi(s) = -\frac{s}{2} \pm\sqrt{\left(A-k+\frac{1}{4}\right)s^2+(k-B)s+C}
\label{eq_1} 
\end{equation}

where A , B and C are defined as
\begin{equation}
A = \epsilon^{2} +\beta_{1} + \beta_{2}; ~~
B = 2\epsilon^{2}+\beta_{1} ; ~~
C = \epsilon^{2}
\end{equation}
The constant k can be determined by setting the discriminate of expression under the square root in eq. \ref{eq_1} to zero, i.e.,$\Delta = b^2 -4ac$ After simple manipulations, we find
\begin{equation}
k_{\pm} = \beta_{1}\pm \epsilon\sqrt{4\beta_2+1}
\end{equation}
Using them, we are able to find four possible solutions for $\pi(s)$ as follows:
\begin{equation}
\pi(s) = \begin{cases} 
\frac{-s}{2} \pm \frac{1}{2}\left[(2\epsilon-\sqrt{4\beta_2+1})s -2\epsilon \right], & \text{for~~k = $k_{+}$}\\ \\
\frac{-s}{2} \pm \frac{1}{2}\left[(2\epsilon+\sqrt{4\beta_2+1})s -2\epsilon \right], & \text{for~~k = $k_{-}$}
\end{cases}
\end{equation}
To make the first derivative of \begin{math} \tau(s) = \tilde{\tau(s)}+2\pi(s) \end{math}
negative, we must select the most suitable form of the $\pi(s)$ as 
\begin{equation}
\pi(s) = \frac{-s}{2} - \frac{1}{2}\left[(2\epsilon+\sqrt{4\beta_2+1})s -2\epsilon \right]
\label{eq2}
\end{equation}
Using eq. \ref{1} and eq. \ref{2}, we can calculate the values of the parameters $\lambda$ and $\lambda_{n}$ as
\begin{equation}
\lambda = \beta_{1}-\frac{1}{2}(1+2\epsilon)(1+\sqrt{4\beta_2+1})
\label{3}
\end{equation}
\begin{equation}
\lambda_{n} = n(1+n+2\epsilon+\sqrt{4\beta_2+1})
\end{equation}
Using the fact that $\lambda = \lambda_{n}$ we obtain the energy
levels as
\begin{equation}
E_{n} = \frac{-\hbar^2\alpha^2}{2\mu}\left[\frac{\beta_1-\frac{1}{4}-(n+\frac{1}{2})[\frac{1}{2}+n+\sqrt{(4\beta_2+1)}}{(2n+1)+\sqrt{(4\beta_2+1)}}\right]^{2}
\end{equation}
For $n=0$, the bound state energy, is given as
\begin{equation}
E = \frac{-\hbar^2\alpha^2}{2\mu}\left[\frac{\beta_1}{1+\sqrt{1+4\beta_2}}-\frac{1}{2}\right]^{2}
\label{gn1}
\end{equation}
After substituting the value of $\beta_{1}$ and $\beta_{2}$ in eq.\ref{gn1}, we get the bound state energy in terms of model parameters as 
\begin{equation}
E = \frac{-\hbar^2\alpha^2}{2\mu}\left[\frac{\frac{2\mu V_1}{\hbar^2\alpha^2}}{1+\sqrt{1+\frac{8\mu V_2}{\hbar^2\alpha^2}}}-\frac{1}{2}\right]^{2}
\label{gnd}
\end{equation}

\subsection{Phase Function Method:} 
The radial time independent Schr$\ddot{o}$dinger equation for $\ell$ orbital angular momentum, is given by
\begin{equation}
\frac{d^2u_{\ell}(k,r)}{dr^2} + \left(k^2-\frac{\ell(\ell+1)}{r^2}\right)u_{\ell}(k,r) = U(r) u_{\ell}(k,r)
\label{SchrEq}
\end{equation}
Where $k=\sqrt{\frac{2\mu E_{cm}}{\hbar^2}}$, $U(r) =\frac{2\mu V(r)}{\hbar^2}$ and $\mu$ is reduced mass of the system.
\\ The $E_{cm}$ is related to $E_{\ell ab}$ by the relation
\begin{math}
    E_{cm} = \frac{m_{T}}{m_{T}+m_{P}}E_{\ell ab} .
\end{math}
Here, $m_{T}$ and $m_{P}$ are masses of target and projectile respectively. 
The second order TISE is transformed into Ricatti type equation\cite{Kynch} which is given by
\begin{equation}
\frac{d\delta_\ell(k,r)}{dr}=-\frac{U(r)}{k}\bigg[\cos(\delta_\ell(k,r))\hat{j}_{\ell}(kr)-\sin(\delta_\ell(k,r))\hat{\eta}_{\ell}(kr)\bigg]^2
\label{PFMeqn}
\end{equation}
where $\hat{j}_{\ell}(kr)$ and $\hat{\eta}_{\ell}(kr)$ are the Ricatti-Bessel and Ricatti-Neumann functions of order $\ell$. 
The phase function got its name because its asymptotic value results in the phase shift, $\delta_{\ell}(\infty) = \delta_{\ell}$.
The phase-function method has the advantage of using a first-order phase equation rather than a second-order Schrödinger equation. The trade-off is that the equation is now nonlinear. Because the angular momentum term is removed in the form given by eq. \ref{PFMeqn}, we must integrate to a greater cutoff point $r_f$ to get adequate convergence for $\delta_{\ell}$ \cite{Xu}. 
For $\ell$ = 0, 1 and 2 (S, P and D waves) respectively, phase equations are: 
\begin{equation}
\frac{d\delta_0(k,r)}{dr} = -\frac{U(r)}{k}\sin^2\left[kr + \delta_{0}(k,r)\right]
\end{equation} 
\begin{equation}
\frac{d\delta_1(k,r)}{dr} =-\frac{U(r)}{k}\bigg[\frac{\sin\left[kr + \delta_1(k,r)\right]-k~ cos\left[kr + \delta_1(k,r)\right]}{kr}\bigg]^2
\end{equation}
\begin{equation}
\frac{d\delta_2(k,r)}{dr} =-\frac{U(r)}{k}\bigg[-\sin{\left[kr + \delta_2(k,r) \right]}-\frac{3 \cos{\left[kr + \delta_2(k,r) \right]}}{kr} + \frac{3 \sin{\left[kr + \delta_2(k,r) \right]}}{(kr)^2}\bigg]^2
\end{equation}
The SPS for S, P and D waves are obtained by numerically solving these equations using 5$^{th}$ order Runge-Kutta method with initial condition chosen as $\delta_\ell(r=0, k) = 0$.
\subsection{Scattering Cross-Section:}
Once, SPS $\delta_{\ell}(E)$ are obtained for each orbital angular momentum $\ell$, one can calculate the partial cross section $\sigma_{\ell}(E)$  using the following formula\cite{Amsler} :
\begin{equation}
\sigma_{\ell} (E) = \frac{4\pi(2\ell+1)}{k^2} \sin^2(\delta_{\ell}(E))
\label{Pxsec}
\end{equation} 
Then, total cross section $\sigma_{T}$, is given as
\begin{equation}
\sigma_{T} = \sigma_{S} +\sigma_{P} +\sigma_{D}
\label{Txsec}
\end{equation}
where $\sigma_{S}$ , $\sigma_{P}$ and $\sigma_{D}$  are given as
\begin{equation}\nonumber
    \sigma_S = \frac{1}{4}\sigma_{^{1} S_0} + \frac{3}{4}\sigma_{^{3} S_1} 
\end{equation}
\begin{equation}\nonumber
    \sigma_P = \frac{3}{12}\sigma_{^{1} P_1} + \frac{1}{12}\sigma_{^{3} P_0} +\frac{3}{12}\sigma_{^{3} P_1} +\frac{5}{12}\sigma_{^{3} P_2} 
\end{equation}
\begin{equation}\nonumber
    \sigma_D = \frac{5}{20}\sigma_{^{1} D_2} + \frac{3}{20}\sigma_{^{3} D_1} +\frac{5}{20}\sigma_{^{3} D_2} +\frac{7}{20}\sigma_{^{3} D_3} 
\end{equation}
\section{Results and Discussion:}\label{sec2}
To optimise the model parameters, we have taken the experimental SPS from Perez et al. of Granada group \cite{Pérez}. They have considered the SPS below pion production threshold up to lab energies of 350 MeV. The model parameters for various channels of S, P and D-states are given in Table \ref{table1}.
\begin{table}[h]
\begin{center}
\caption{Model parameters of Deng-Fan Potential for various partial waves of S, P and D states of np scattering.}
\begin{tabular}{@{}rrrr@{}}
\hline
States & $V_{1} (MeV)$      & $ V_{2} (MeV)$       & $\alpha (fm^{-1})$   \\ \hline
$^{3}S_{1}$   &  728.3840  & 1443.0481  & 2.0429  \\ 
$^{1}S_{0}$    &450.3160  & 918.3518 &1.9291  \\ \hline
$^{1}P_{1}$     & 0.4147  & 43.7142   & 0.6111 \\
$^{3}P_{0}$     & 282.3228  & 1722.3388  & 1.5338 \\
$^{3}P_{1}$     & 0.4147    & 219.1505   & 0.9784 \\
$^{3}P_{2}$     & 409.5267  & 521.4924  & 2.4215 \\ \hline
$^{1}D_{2}$     & 189.2142  & 0.4147     & 1.7670  \\
$^{3}D_{1}$    & 0.4147   & 59.7035   & 0.5073 \\
$^{3}D_{2}$     & 259.5831  & 253.8225   & 1.3185 \\
$^{3}D_{3}$     & 1237.8333 & 18665.3004 & 2.9657 \\ \hline

\label{table1}
\end{tabular}
\end{center}
\end{table}

\begin{figure}[hbtp]
\centering
\includegraphics[scale=0.35]{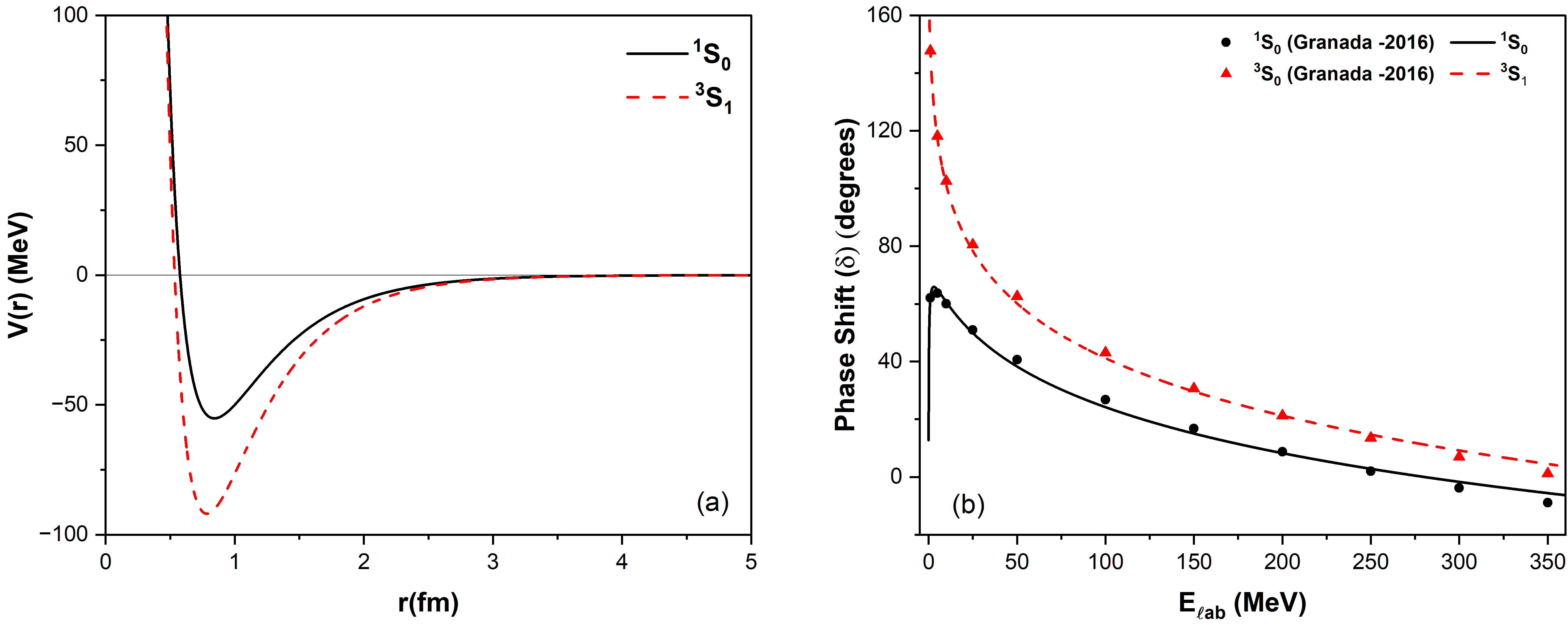}
\caption{(a) $^3S_1$ and $^1S_0$ potentials and (b) Corresponding scattering phase shifts}
\label{Fig1}
\end{figure}

For $^3S_1$ ground state, we have utilised the energy condition given by eq. \ref{gnd} and substituted $E = -2.224549~MeV$ for binding energy of deuteron\cite{Breit}. So, choosing parameters $V_{2}$ and $\alpha$ in Deng- fan potential, one can obtain $V_1$ as 
\begin{equation}
V_1 = \frac{\hbar^2\alpha^2}{2\mu}\left[1+\sqrt{1+\frac{8\mu V_2}{\hbar^2\alpha^2}}~\right]~\left[\sqrt{\frac{-2\mu E}{\hbar^2\alpha^2}}+\frac{1}{2}~\right]
\end{equation}
Only two parameters need to be optimised for obtaining corresponding interaction potential. Hence, it is possible to incorporate the energy condition while determining SPS using PFM. This is contrary to one of the conclusions drawn by Saha et.al.,\cite{Saha}, where they claim that Jost function method (JFM) has supremacy over PFM due to the fact that the parameters of the potential are determined by ﬁtting proper binding energies for the system. Further, they also claim that PFM gives little discrepancies in SPS as compared to JFM. But, our phase-wave analysis using PFM for both $^3S_1$ and $^1S_0$ states matches expected data\cite{Pérez} for not only upto 50 MeV but all the way up to 350 MeV. The obtained interaction potentials are shown in Fig. \ref{Fig1}(a) and corresponding SPS for S-states are plotted along with data taken from Perez et al., \cite{Pérez} in Fig. \ref{Fig1}(b).
\\ The low-energy scattering parameters, namely the scattering length ($a$) and effective range ($r$), are determined for the $^1S_0$ and $^3S_1$ states using the effective-range approximation formula \cite{Babenko}.
For $^1S_0$ state, the scattering length is found to be $-24.016 ~fm $ ($-23.749(8) ~fm$) and the effective range is $2.49~ fm$ ($2.81(5)$ fm).
For $^3S_1$ state, the scattering length is determined to be $5.477 ~ fm$ ($5.424(3) ~fm$), while the effective range is $1.89 ~fm$ ($1.76(5)$ fm).
%
One can observe from Fig.\ref{Fig1}(a) that potentials for triplet and singlet states look similar except for their depth of interaction, which is expected, due to different contributions from their spin-spin interactions. The obtained SPS are well matched with expected ones upto 250 MeV, but at 300 MeV and 350 MeV there is a discrepancy of 2$^{o}$ and 3$^{o}$ for both $^{1}S_{0}$ and $^{3}S_{1}$ state.\\
\begin{figure}[hbtp]
\centering
\includegraphics[scale=0.35]{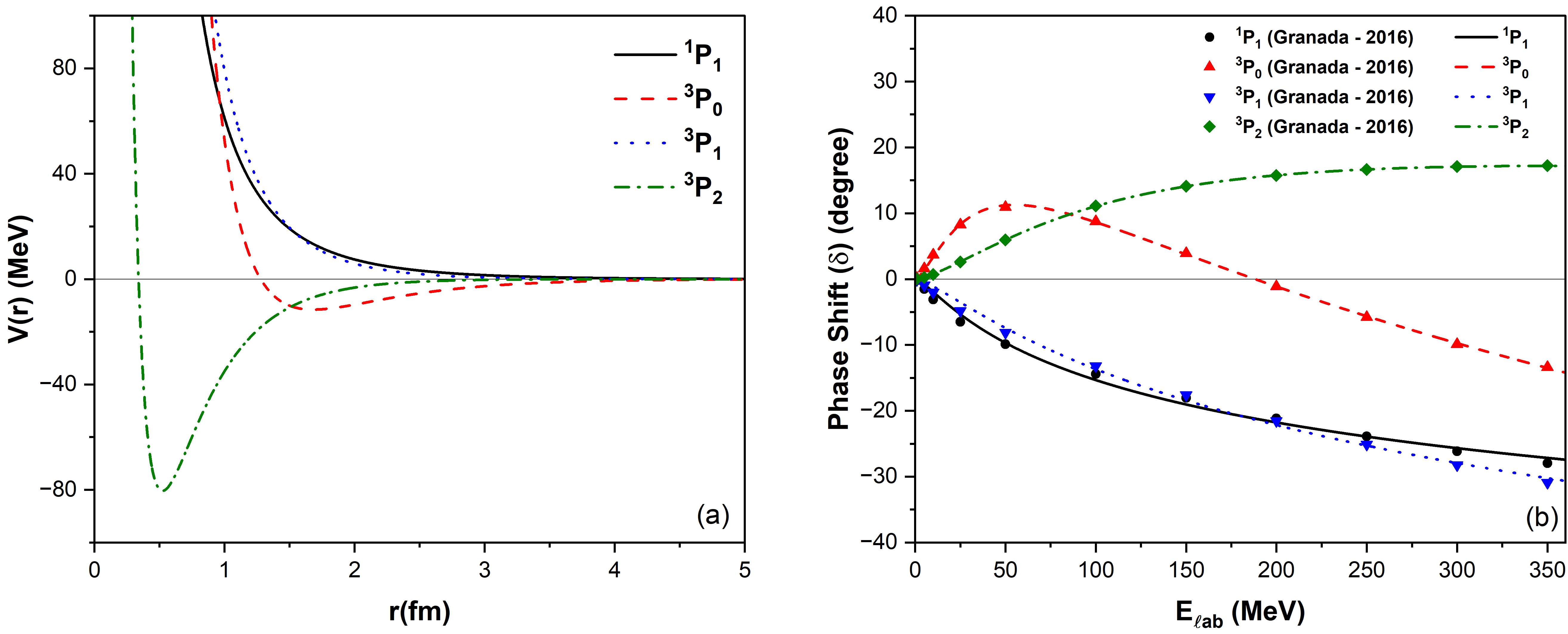}
\caption{(a) P-wave interaction potentials and (b) Corresponding scattering phase shifts}
\label{Fig2}
\end{figure}

\begin{figure}[hbtp]
\centering
\includegraphics[scale=0.35]{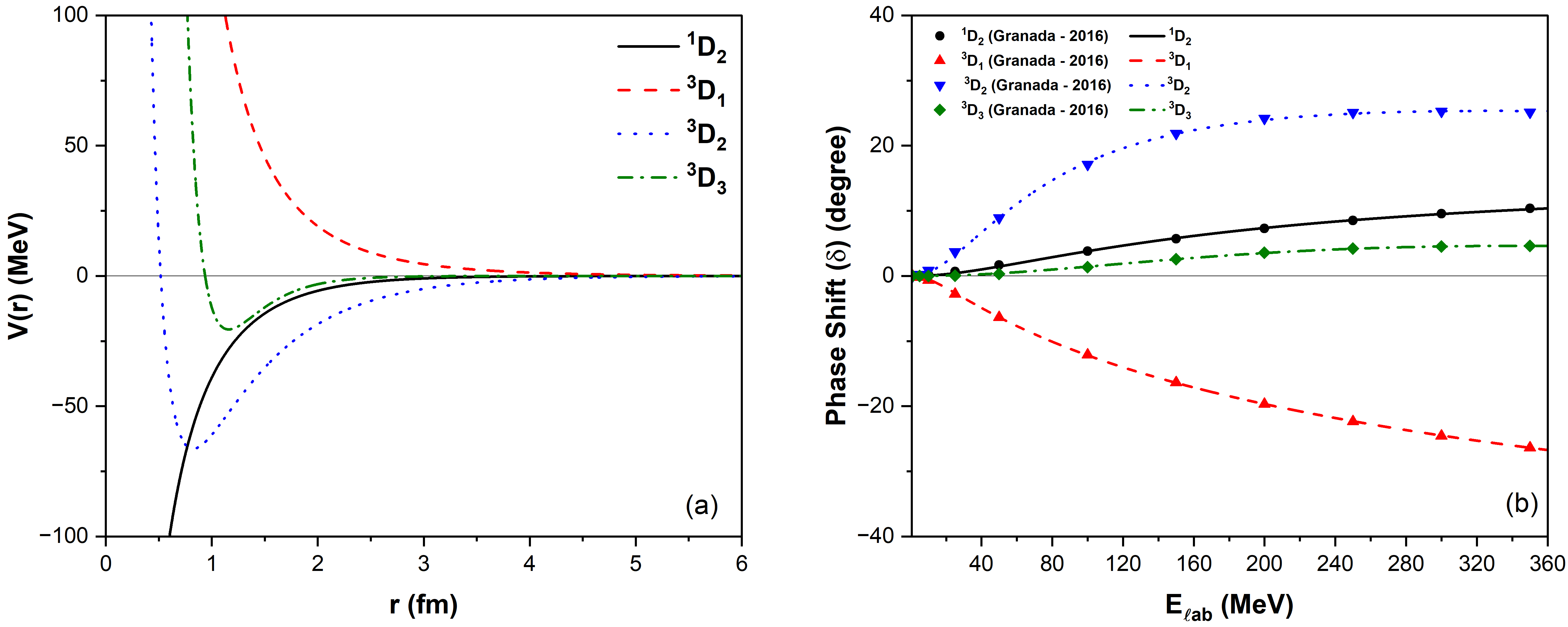}
\caption{(a) D-wave interaction potentials (b) Corresponding scattering phase shifts}
\label{Fig3}
\end{figure}
\begin{figure}[h]
    \centering
    \includegraphics[scale=0.35]{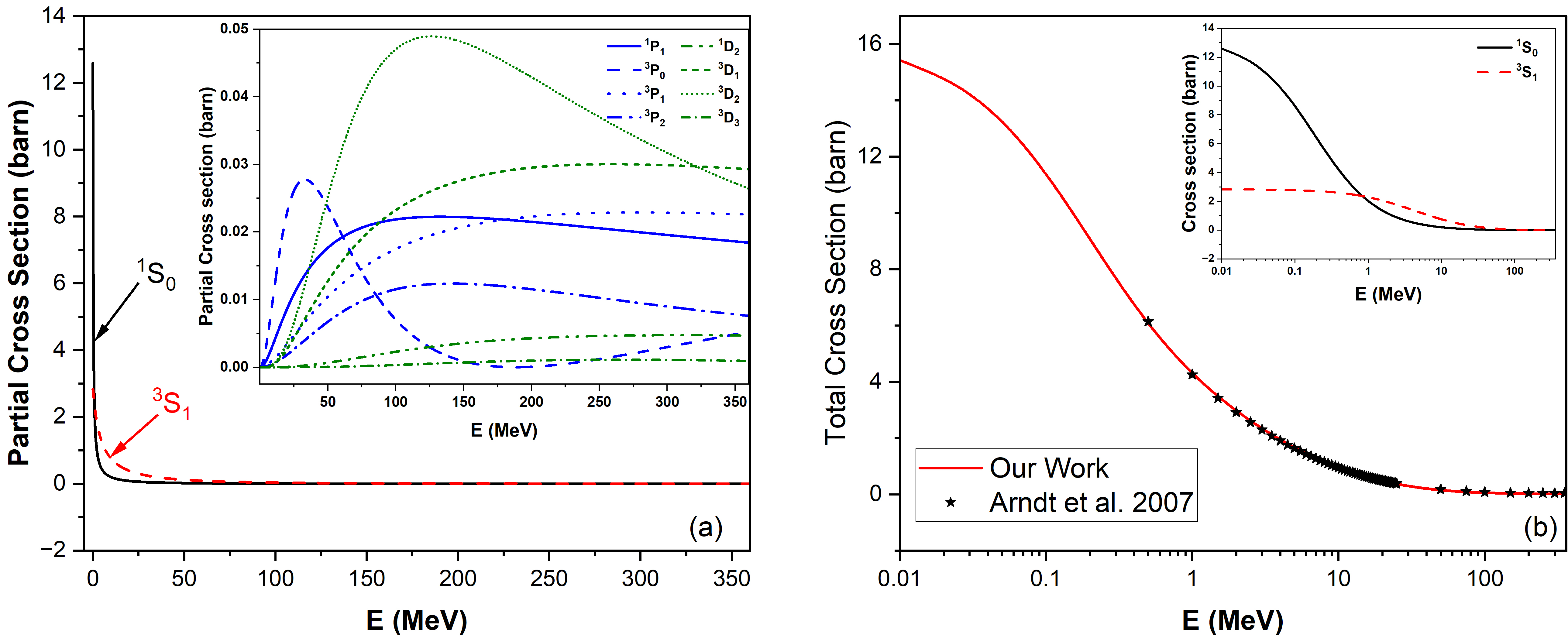}
    \caption{(a) Partial scattering cross-sections for Singlet and Triplet S-state and those of P and D-states are shown in inset (b) Total elastic scattering cross-section for np interaction.}
    \label{fig4}
\end{figure}

The methodology is then extended to study SPS for P and D waves wherein one expects spin-orbit interaction to play an important role. The centrifugal terms for P and D states are taken care of in respective phase equations for $\ell = 1$ and $\ell = 2$ in PFM. Typically, spin-orbit term is modelled as derivative of central potential. Since, DF potential is basically combination of exponential terms, its derivative would also result in a more complicated combination of exponential terms and with larger powers for expressions in their denominators. Certainly, one way of determining model parameters would be to simultaneously optimise them to fit expected SPS data for all channels. That would increase the total number of parameters to be simultaneously optimised and hence the computational cost. Actually, one is interested only in interaction potential, responsible for observed SPS, for each of the channels. This would basically be obtained by substituting overall model parameters in a potential consisting of various contributions from central, spin, spin-orbit, etc. If the same can be achieved by refitting model parameters of the phenomenological potential, there is no loss in information. This procedure of obtaining model parameters  has been undertaken using PFM while studying np, pp, n$\alpha$ and p$\alpha$ systems already\cite{Behera,Kumar,Kumar_1,Arushi}. The overall potentials, that include all contributions of underlying interactions, for various P and D states are shown in Figs. \ref{Fig2}(a) and \ref{Fig3}(a) respectively. The obtained SPS, for data up to 350 MeV, are shown in Figs. \ref{Fig2}(b) and \ref{Fig3}(b) respectively. The observed match between SPS obtained using PFM and expected SPS\cite{Pérez}, very much confirm the points raised in the above discussion.\\
\begin{table}[h]
\begin{center}
\caption{The differential and total elastic scattering cross-section(SCS):\\ The $\%$-contribution of each channel to experimental SCS is given in brackets next to differential SCS}
\scalebox{0.8}{
\setlength{\tabcolsep}{12pt} 
\renewcommand{\arraystretch}{1.5}
\begin{tabular}{@{}lllllll@{}}
\hline
E & $\sigma_{exp}$(b)  & $\sigma_{^{1}S_{0}}$  &$\sigma_{^{3}S_{1}}$ & $\sigma_{P}$ & $\sigma_{D}$ & $\sigma_{sim}$(b) \\ 
(MeV) & \cite{Arndt} & & & & & \\ \hline
0.1 & -       & 8.610 (76\%)        & 2.772 (24\%)         & 3.75013$\times 10^{-7} $   & 2.5$\times 10^{-14} $       & 11.382         \\
0.5 & 6.135   & 3.551 (58\%)   & 2.532 (42\%)   & 9.22926$\times 10^{-6} $     & 1.57947$ \times 10^{-11} $    & 6.083   \\
1   & 4.253   & 2.029 (47\%)  & 2.282 (53\%)   & 3.61667$\times 10^{-5} $     & 2.45356$\times10 ^{-10} $    & 4.311    \\
10  & 0.9455  & 0.1980 (21\%)  & 0.7530 (79\%)  & 0.0023         & 1.71628$\times 10 ^{-6} $    & 0.9533   \\
50  & 0.1684  & 0.0200 (13\%) & 0.1177 (78\%)  & 0.0118 (7\%)   & 0.0004        & 0.1499  \\
100 & 0.07553 & 0.00438 (8\%) & 0.03386 (62\%) & 0.01523 (28\%) & 0.00151 (2\%) & 0.05497\\
150 & 0.05224 & 0.00117 (4\%)  & 0.01273 (40\%)   & 0.01597 (50\%) & 0.00213 (6\%) & 0.03200 \\
200 & 0.04304 & 0.00027 (1\%) & 0.00514 (22\%) & 0.01576 (67\%) & 0.00236 (10\%) & 0.02352 \\
250 & 0.03835 & 0.00003 & 0.00200 (10\%)  & 0.01522 (78\%) & 0.00238 (12\%) & 0.01963 \\
300 & 0.03561 & 0.00001   & 0.00066 (4\%)  & 0.0146 (83\%)  & 0.00227 (13\%) & 0.01754 \\
350 & 0.03411 & 0.00007 & 0.00014 (1\%)        & 0.01397 (86\%) & 0.00211 (13\%) & 0.01629 \\ \hline
\label{table2}
\end{tabular}
}
\end{center}
\end{table}
The partial and total scattering cross-sections are obtained using eq.      \ref{Pxsec} and eq.\ref{Txsec} respectively. The individual contributions due to $^1 S_0 $ and $^3 S_1 $ and overall contribution due to P and D waves are given in table \ref{table2}. One can observe that the contributions from P and D states become comparable for higher energies. It is seen that the discrepancies between experimental and observed SCS increases with increasing energy. The differential cross section for both states of S wave are plotted in Fig. \ref{fig4}(a), with those of P and D states as inset. The total cross section plot with logarithmic energy scale is shown in Fig. \ref{fig4}(b), with an inset of contributions from $^3S_1$ and $^1S_0$.\\ 
In Fig. \ref{fig4}(a), it is observed that, contribution from $^1S_0$ state is much larger than the  $^3 S_1$ state. The contribution from P and D waves are very less as compared to those from S waves at low energies. The obtained total cross sections are very well matched with experimental cross sections as shown in Fig. \ref{fig4}(b). In its inset, one can observe that beyond 1 MeV, $^3S_1$ has greater contribution to total scattering cross-section than $^1S_0$. This is because, while the scattering state has energy close to zero, about 77 keV, the ground state has energy of 2.2245 MeV.\\ 
It would be interesting to see the performance of Deng-Fan potential by considering all other higher $\ell$-channels of np-interaction. It is also important to cross-check it's effectiveness in explaining the experimentally observed Deuteron properties. In this paper, we have limited the scope of study to only understand np-scattering through interaction modeled using DF potential and obtained total cross-sections to validate its effectiveness, in explaining experimentally observed cross section.

\section{Conclusion}\label{sec13}
The Deng-Fan potential, which is a combination of attractive Hulthen potential and a repulsive part which is square of the Hulthen term, has the advantage of having analytical solutions for time independent Schrodinger equation. Being a combination of Hulthen terms, it should have been utilised as model of interaction for understanding np scattering. This has been achieved for S-waves for lab energies up to 50 MeV using Jost function method and parallely using phase function method \cite{Saha}. In this work, we have extended the phase wave analysis, for lab energies up to 350 MeV, by obtaining scattering phase shifts for not only S-waves but also P and D waves. The total scattering cross-sections have been obtained by determining partial cross-sections for each of the S, P and D states and are shown to match very closely with experimental ones over the entire range of energies. Hence, one can conclude that Deng-Fan potential is a suitable phenomenological potential to study np-interaction. It would be interesting to see its performance in studying other scattering systems such as n-D, p-D, n-$\alpha$, p-$\alpha$, $\alpha-^{3}He$, $\alpha-^{3}H$ etc.

\textbf{Acknowledgments}
\\
A. Awasthi acknowledges financial support provided by Department of Science and Technology
(DST), Government of India vide Grant No. DST/INSPIRE Fellowship/2020/IF200538. The authors dedicate this effort to memory of Late Prof. H.S. Hans, during his birth centenary celebrations.
 
\end{document}